\documentclass[twocolumn,prb,aps,showpacs,footinbib,amsmath,amssymb,superscriptaddress]{revtex4-1}
\usepackage{graphicx}
\usepackage{dcolumn}
\usepackage{color}
\usepackage{bm}
\usepackage{natbib,hyperref}
\usepackage{amsmath,amssymb,amsfonts,bbold}
\usepackage[normalem]{ulem}

\begin{document}

%The next command sets up an environment for the abstract to your paper.

\title{Calculations of in-gap states of ferromagnetic spin chains on \textit{s}-wave wide-band superconductors}

\author{Cristina Mier}
\affiliation{Centro de F{\'{\i}}sica de Materiales
        CFM/MPC (CSIC-UPV/EHU),  20018 Donostia-San Sebasti\'an, Spain}
\author{Deung-Jang Choi}
\affiliation{Centro de F{\'{\i}}sica de Materiales
        CFM/MPC (CSIC-UPV/EHU),  20018 Donostia-San Sebasti\'an, Spain}
\affiliation{Donostia International Physics Center (DIPC),  20018 Donostia-San Sebasti\'an, Spain}
\affiliation{Ikerbasque, Basque Foundation for Science, 48013 Bilbao, Spain}
\author{Nicol{\'a}s Lorente}
\email{nicolas.lorente@ehu.eus}
\affiliation{Centro de F{\'{\i}}sica de Materiales
        CFM/MPC (CSIC-UPV/EHU),  20018 Donostia-San Sebasti\'an, Spain}
\affiliation{Donostia International Physics Center (DIPC),  20018 Donostia-San Sebasti\'an, Spain}

\begin{abstract}
Magnetic impurities create in-gap states on superconductors. 
Recent experiments explore the topological properties 
of one-dimensional arrays of magnetic impurities
on superconductors, because in certain regimes p-wave pairing can
be locally induced leading to new topological phases. A by-product
of the new accessible phases is the appearance of zero-energy edge states
that have non-Abelian exchange properties and can be
used for topological quantum computation. 
Despite the large amount of theory devoted to these systems,
most treatments use approximations that render their applicability
limited when comparing with usual experiments of 1-D impurity arrays
on wide-band superconductors. These approximations either
involve tight-binding-like approximations where the impurity
energy scales match the minute energy scale of the superconducting
gap and are many times unrealistic, or they assume strongly-bound
in-gap states. Here, we present a theory for s-wave
superconductors based on a wide-band normal metal, with any possible energy scale for
the magnetic impurities. The theory is based on free-electron
Green's functions. We include
Rashba coupling and compare with recent experimental
results, permitting us to analyze the topological phases
and the experimental edge states. The infinite-chain properties can be analytically obtained, giving us a way to compare with finite-chain calculations. We show that it is possible to converge to the infinite limit by doing finite numerical calculation, paving the way for numerical calculations not based on analytical Green's functions. 
\end{abstract}

\date{\today}


\maketitle

\section{Introduction}
Ferromagnetic spin chains on s-wave superconductors have been shown to be able to
host Majorana bound states (MBS) at the edges of the chains by both theoretical
~\cite{Beenakker,IvarMartin,Yazdani_2013,Brenevig,Pientka2013,Simon1,Ojanen}
and experimental works~\cite{Yazdani,Pawlak_2016,Yazdani2017,Kim_2018,Choi_2019}.
Atomic magnetic impurities produce in-gap states~\cite{Yu,Shiba,Rusinov}
because the local magnetic interaction weakens the binding of Cooper
pairs permitting the presence
of one-particle states in the superconductor gap~\cite{Yazdani_1997,Flatte_1997a,Flatte_1997b}. When several
impurities lie on the surface such that their induced in-gap states
can overlap, in-gap bands can be formed. For two impurities aligning
ferromagnetically, the in-gap states form delocalized-state that are analogues of
molecular orbitals~\cite{Flatte_2000,Morr_2006,Yao2014} in a one-electron
picture of molecular binding.  The characterization of molecular-like
in-gap states has been made possible by the scanning tunneling
microscope (STM)~\cite{Liljeroth_2018,Choi_2018,Beck_2020,Ding2021}. As more impurities are added,
delocalized states form in ferromagnetic structures such that their lowest
eigenergies can cross
the chemical potential level of the superconductor
as has been found experimentally~\cite{Schneider1,Mier2021b}.
Thus, the spin chain can lead to closing of the superconducting
gap and to a phase transition. In the presence of spin-orbit
coupling, the pairing can change and a topological phase transition (TPT) is induced~\cite{Brydon,heimes_2015}. A helical spin
structure~\cite{IvarMartin,Simon1,Simon2,Pientka2013,Ojanen,Paaske2016} 
has been shown
to be equivalent to the effect of a Rashba interaction caused by the
superconductors spin-orbit coupling~\cite{Pientka2015,Paaske2016,Paaske1}. The possibility of emergence of MBS in antiferromagnetically ordered chains has been also discussed\cite{Kotetes, Kobialka}.\\
Recent experiments show that ferromagnetic spin chains can be assembled
on s-wave superconductors that have a substantial Rashba spin-orbit coupling~\cite{Yazdani,Pawlak_2016,Yazdani2017,Kim_2018,Schneider1,Mier2021b,Schneider2}. These
structures can create topological phases that should show MBS at the edges of the structures even in the presence of substantial disorder~\cite{Awoga}.
Usual s-wave superconductors are normal metals above the superconducting
transition, with large bands. Typical parameters show that the superconducting
gap is several thousands of time smaller than the superconductor's band. This situation
is very different from semiconductor nanowire systems proximitized by an s-wave superconductor~\cite{vonOppen,dasSarma,Aguado}
where the induced superconductor gap and the band gap can be expected
to be of the same order of magnitude. The effect of magnetic impurities
 is usually accounted for by an exchange interaction
 acting on the superconductor's electrons. Realistic exchange interactions on solids range from hundreds of meV to eV, while
exchange fields in semiconducting nanowires only close the gap when they are
of the same magnitude as the superconductor gap. The large difference in energy scales between impurities on wide-band superconductors
and proximitized semiconducting nanowires lead to very different techniques to treat each case.

The usual approach to treat
wide-band superconductors is to approximate
the metallic phase by a free-electron metal \cite{Flatte_1997a,Flatte_1997b,Pientka2013,Paaske2016,Bernevig,Choi_2018,Schneider1,Schneider2}. The resulting
superconductor comes after a series of approximations to have
the expected real-space properties such as Friedel-like
oscillations and coherence lengths in the range of hundreds of nanometers. The following
step is either to maintain a Green's function approach to characterize the electronic properties
of the superconductor, and include magnetic impurities via Dyson's equation~\cite{Flatte_1997a,Flatte_1997b,Choi_2018} (or equivalently the T-matrix~\cite{Paaske2016,CristinaBena}),
or to solve for the wave functions using a Lippmann-Schwinger approach~\cite{Pientka2013,Bernevig,Schneider1,Schneider2}.
In the presence of magnetic impurities and helical spin ordering, the wave-function approach is further simplified
by assuming that the induced in-gap states are strongly bound and hence close to the Fermi energy.
Despite this last approximation, both approaches are very similar because they include the
same approximations to treat the free-electron-based superconductor.

The Green's function approach is the one used in the present work. In section~\ref{Green}, we will briefly show
the main approximations and features of the method. In particular, it has the advantage of
a simple and efficient numerical implementation where the main operations are matrix
inversions. Previous works have shown that this technique
permits us to rationalize the
experimental findings in wide-band superconductors with large impurity exchange coupling~\cite{Choi_2018,Mier2021b}.
The imaginary part of the real-space Green's function is the local density of states (LDOS), then
using Tersoff-Hamman's theory \cite{Tersoff-Hamann}, we can evaluate the quantities obtained in STM experiments by real-space Green's functions. The study
of topological phases can be evaluated by determining the presence of MBS in the LDOS.
However, as in experiments, it can be  difficult
to prove that a zero-energy edge mode is a MBS. 
There are two usual ways of identifying MBS. One is to study the
properties of the MBS.
Previous works have shown that the spin-polarization of MBS are a crucial
quantity to determine~\cite{Sticlet,Yazdani2017,Mahdi2020,WangJens}.
As shown in Ref.~[\onlinecite{Mahdi2020}], the 
study of the evolution
of the in-gap-band structure with the changing parameter
together with the spin polarization is an excellent
probe to unravel topological phases. Comparison of the Green's function
obtained LDOS with
experiments indeed show that Cr spin chains on $\beta$-Bi$_2$Pd wide-band superconductor
can undergo a topological phase transition when the number of atoms increase~\cite{Mier2021b}.
There, it is also shown that the staggered magnetic moment gives valuable information
as predicted in Ref.~[\onlinecite{Sticlet}].
A second way is to use the bulk-boundary correspondence principle\cite{Asboth,Kitaev_2001}, where the study of topological invariants determined by the bulk Hamiltonian implies the appearance of MBS.

In this article, we have chosen the second approach.
In section~\ref{Shiba}, we will describe the impurity and Rashba Hamiltonians used here, and the derivation of the in-gap-bands. To this end, we implement the theory developed by Tewari and Sau~\cite{Tewari2012} using Green's functions.
We find that the usual implementation~\cite{WangZhang} identifying
the reciprocal space Hamiltonian with the zero-energy
inverse of the Green's function does not work because the superconducting
real-space Green's function is not a resolvent. Instead, we derive an effective Hamiltonian that we obtain from the renormalization of the Green's function and that correctly describes the dispersion of the in-gap states. From this band structure, in section \ref{IV}, we compute the winding-number and the lower-symmetry $\mathbb{Z}_2$ invariant for an infinite 1-D system. These quantities determine the topological phase of the system, in good agreement with the emergence of MBS at the edge of finite chains. A systematic study of ferromagnetic spin chains allows us to define topological phase diagrams. In section \ref{V}, we compare these results with numerical calculations on finite systems. Our results show that winding numbers can still be calculated in these systems. Indeed, our numerical procedure seems to be robust and can be used to analyze 1-D chains in 2-D superconductors with increasing resemblance to the experiment. As an example, we are able to identify the non-topological or topological origin of edge modes that oscillates around the Fermi energy as atoms are added to the chain. Similar behavior has been recently reported on experimental results\cite{Schneider1}.

\section{Wide-band electronic Green's function for superconductors}
\label{Green}
Local-basis-set approaches to treat a wide-band superconductor incurs
into numerical problems due to the large energy mismatch between
the electron-band width and the pairing energy $\Delta$. Here, we will
use extended states to describe the superconductor. The normal metal is
then treated in the free-electron approximation. We adopt the
theory developed by Flatt\'e and Byers~\cite{Flatte_1997a,Flatte_1997b}.
The main features of their theory is to use a real-space
Green's function for the superconductor based on free-electrons and
then solve for the effect of magnetic impurities
using Dyson's equation. In this section, we are going to briefly analyze
the specificities of the Green's functions obtained in this way.

The Bogoliubov-de Gennes approach can be succinctly expressed
using Nambu's formalism~\cite{Shiba,Balatsky_2006,zhu2016,Vernier,Pientka2013,Paaske2016}.
Here, we choose a plane-wave electronic basis and 4-component Nambu operators
to express the Bogoliubov-de Gennes equations in matrix form.
The basis set of our study is given by the Nambu operator~\cite{zhu2016,Vernier}
$\hat{\Psi}_k = (\hat{c}_{k,\uparrow}, \hat{c}_{k,\downarrow}, \hat{c}_{-k,\uparrow}^\dagger, \hat{c}_{-k,\downarrow}^\dagger)^T$,
where, $k$ is the wave-vector of the plane-wave basis function $\phi_k(r) =  e^{i \vec{k} \cdot \vec{r}}/\sqrt{V}$,  $V$ is the normalization volume and $\vec{r}$ are the spatial-coordinate vectors.

The Bogoliubov-de Gennes formalism is a mean-field treatment that permits
us to use one-particle equations at the expense of doubling the basis set. Indeed,
the Hamiltonian resulting from using the Nambu's basis set is artificially particle-hole
symmetric and contains two extra redundant solutions that are not physical.
In many cases, the basis set can be reduced to a 2-component Nambu spinor. This
is not possible when spin-flip scattering or spin-orbit interactions
are present due to the mixing of spinor components~\cite{zhu2016}.

Using free-electrons with a pairing interaction of strength $\Delta$, the Hamiltonian
matrix is expressed as~\cite{Shiba,Vernier},
\begin{equation}
\label{BCS}
{H}_{BCS} = \xi_k\tau_z\sigma_0 + \Delta \tau_y\sigma_y,
\end{equation}
where $\xi_k$ is the energy from the Fermi level ($\xi_k = \epsilon_k - E_F$), and $\Delta$ is the superconducting pairing potential.
Here, the tensor product of Pauli matrices for the spin ($\sigma$) and particle ($\tau$) sectors  spans
the $4 \times 4$-matrix space if the identity matrices ($\sigma_0$ and $\tau_0$) are included.
Thanks to the one-particle character of this Hamiltonian, we can use the resolvent to find the
superconductor's Green's function~\cite{Vernier}:
\begin{eqnarray}
\label{GoBCS}
G^0(k,\omega) &=& \langle k | [\omega \tau_0 \sigma_0-{H}_{BCS}]^{-1} | k \rangle \nonumber \\ 
&=& 
\frac{1}{\omega^2 - \xi^2 - \Delta^2} (\omega \tau_0 \sigma_0 + \xi_k\tau_z\sigma_0 + \Delta \tau_y\sigma_y).
\end{eqnarray}
The retarded version of Eq. (\ref{GoBCS})
can be obtained replacing $\omega$ by $\omega+ i \Gamma$, where the Dynes parameter, $\Gamma$, is taken
as a small and positive real number that is phenomenologically associated
with the lifetime of quasiparticles \cite{Dynes}. The imaginary part of $-G^0(k,\omega)$
becomes the one-particle density of states of the superconductor. We do not even attempt to plot
this density of states for a realistic wide-band superconductor because it basically
reduces to two tiny gaps with value $2\Delta$ near the Fermi wave vector $k_F$ that disappear in the fast dispersing
bands with $k$-values.

Instead, we will Fourier transform to real space. However, the Fourier transform does not converge.
We are forced to apply the BCS-like trick of only including states within a shell of width
the Debye energy, $\hbar \omega_D$,
around the Fermi energy~\cite{Schrieffer1994}. This can be done weighing the integrand by
Gaussian functions of width  $\hbar \omega_D$, and at the end of the calculation taking
the limit  $\hbar \omega_D \rightarrow 0$ because the Debye energy is much smaller than the Fermi
energy. This is done in Ref.~[\onlinecite{Pientka2013}]. This set of approximations allows us
to recover the real-space Green's function for a free-electron-like superconductor as
presented by Flatt\'e and Byers~\cite{Flatte_1997a,Flatte_1997b}.
In the above $4 \times 4$-Nambu space, the non-local Green's function is:
\begin{widetext}
\begin{eqnarray}
\label{GBCS}
  &{G}_{BCS}&(r,\omega) = -\frac{\pi N_0}{k_Fr}e^{\frac{-\sqrt{\Delta^2 - \omega^2}}{\pi\xi\Delta}r}  \\
&\times&
\begin{pmatrix} 
    \cos{k_Fr} + \frac{\omega}{\sqrt{\Delta^2 - \omega^2}}\sin{k_Fr} & 0 & 0 & \frac{-\Delta}{\sqrt{\Delta^2 - \omega^2}}\sin{k_Fr} \\
    0 & \cos{k_Fr} + \frac{\omega}{\sqrt{\Delta^2 - \omega^2}}\sin{k_Fr} & \frac{\Delta}{\sqrt{\Delta^2 - \omega^2}}\sin{k_Fr} & 0 \\
    0 & \frac{\Delta}{\sqrt{\Delta^2 - \omega^2}}\sin{k_Fr} & -\cos{k_Fr} + \frac{\omega}{\sqrt{\Delta^2 - \omega^2}}\sin{k_Fr} & 0 \\
    \frac{-\Delta}{\sqrt{\Delta^2 - \omega^2}}\sin{k_Fr} & 0 & 0 & -\cos{k_Fr} + \frac{\omega}{\sqrt{\Delta^2 - \omega^2}}\sin{k_Fr} \nonumber
\end{pmatrix}
\end{eqnarray}
\end{widetext}
where $r$ is the distance between two points in the superconductor. The prefactor includes $N_0$ that is the normal-metal density of states at the Fermi energy, and the exponential
behavior with distance, controlled by the correlation length $\xi$ of the superconductor.
This expression recovers known properties of the electronic structure of
an s-wave superconductor. However, it presents a divergence at $r=0$. We can just evaluate the Fourier
transform for $r=0$ and find out the correct expression \cite{Vernier,Paaske2016} of the Green's
function at $r=0$. In this way, we will have the correct expression for finite $r$ and
 the correct $r=0$ limit. This should not be a problem when using the above expressions
on a lattice, such that the discrete step is of the order of
the underlying lattice parameter. Since the spatial oscillations appearing in Eq. (\ref{GBCS})
are of the order of the Fermi wavelength, $ \lambda_F = 2 \pi/k_F$, we will be able
to use Eq. (\ref{GBCS}) in a discrete lattice~\cite{Meng} such that $\Delta r > 1/k_F$. In practice, we
can probably use Eq. (\ref{GBCS}) even for rather small values of $k_F$ and
still have physically-correct results down to steps of the order of the lattice parameter.
Following Ref.~[\onlinecite{Vernier}], the $r=0$ limit of the  Green's function is
the usual local BCS Green's function:
\begin{eqnarray}
G_{BCS} (r=0, \omega)
&=& - \frac{\pi N_0 Sgn [Re(\omega)Im(\omega)]}{\sqrt{\Delta^2-\omega^2}} \nonumber \\
&\times&
\begin{pmatrix}
\omega & 0 & 0 & -\Delta \\
0 & \omega &\Delta& 0\\
0 &\Delta&\omega& 0 \\
-\Delta & 0 & 0 & \omega
\end{pmatrix}.
\end{eqnarray}

In summary, the above real space Green's function has two important restrictions. The
first one is that beyond an energy scale given by the Debye frequency, the
Green's function is not physical. The second one is that it
can be used seamlessly from $r=0$ to finite $r$ if the discrete steps are large enough,
where the typical length scale is given by the Fermi wavelength.

In the present work, we are interested in studying the topological phases
associated with spin chains on wide-band superconductors. We will evaluate
the topological properties of the bulk superconductor. To do this, we need to
transform back our real space Green's function to k-space. As in previous
works~\cite{Pientka2013,Paaske2016}, we are going to assume a discrete spatial
step $\Delta r= a$, the lattice parameter of our superconductor. In this
case, we have to evaluate the discrete Fourier transform that
due to the translational invariance of the underlying crystal structure
can be written as~\cite{Asboth}
\begin{equation}
        G_{BCS} (\vec{k}, \omega)= \sum_{\vec{R}} 
        {G}_{BCS} (R,\omega) e^{i \vec{k} \cdot \vec{R}}.
        \label{GBCSr}
\end{equation}
Here $\vec{R}$ are all the positions of the atoms in the crystal. We
will work on 1-D spin chains. Then it is interesting to find $G_{BCS} (k, \omega)$
in 1-D where the other two spatial coordinates have been set to zero.
This is easily done because the sum over $\vec{R}$ can be analytically
performed as explained in Ref.~[\onlinecite{Pientka2013}]. For
the Green's function this has been done in the supplemental material
of Ref.~[\onlinecite{Paaske2016}].

\begin{figure}[t]
\begin{center}
\includegraphics[width=0.35\textwidth]{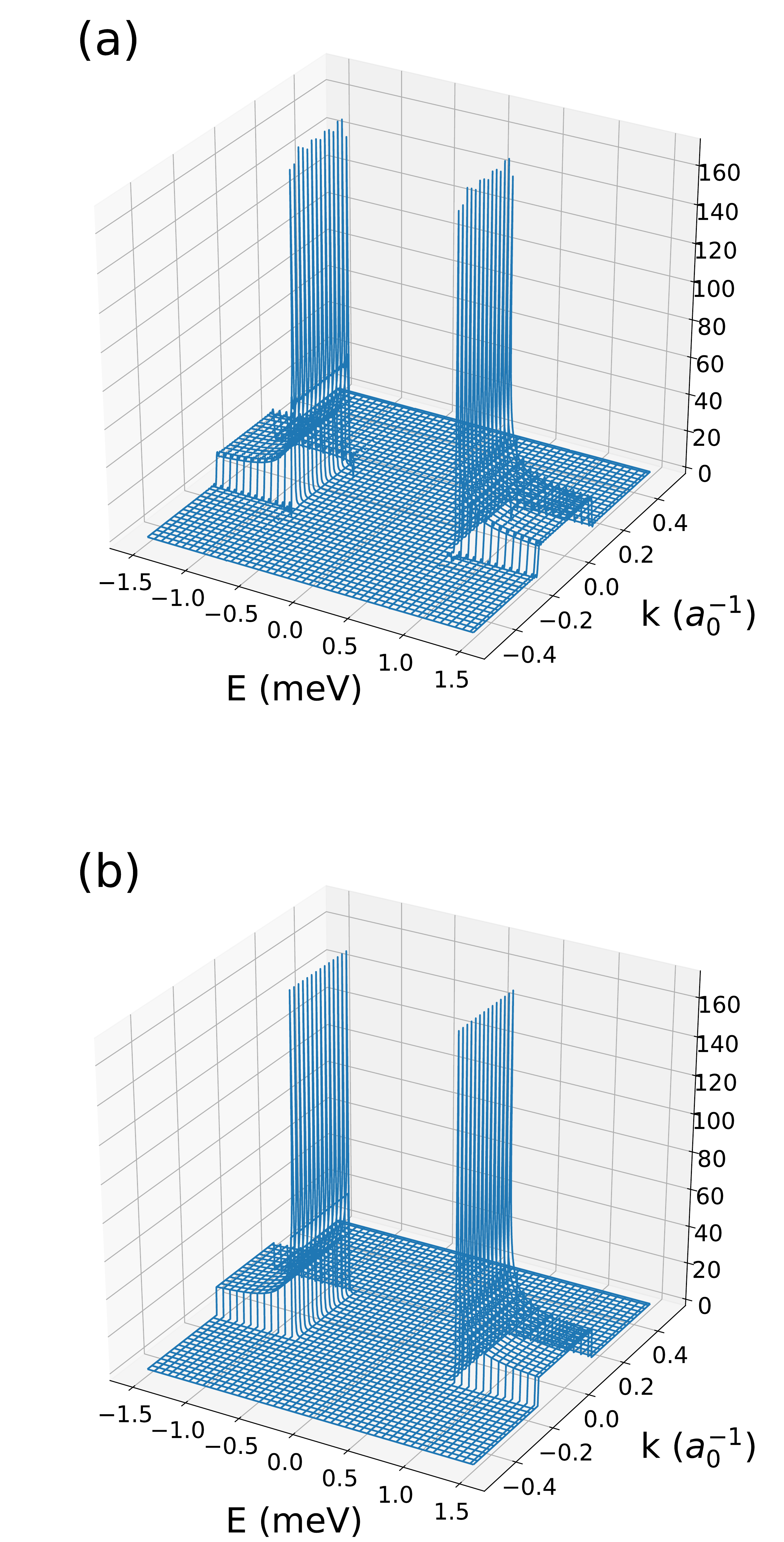}
\end{center}
        \vspace{-0.3cm}
        \caption{
Density of states for a free-electron-like
superconductor computed using Eq.~(\ref{GBCSr}), for $k_F=0.15\; a_0^{-1}$ and  $\pi/a = 0.5\; a_0^{-1}$.
 $(a)$ is the calculation using a finite set of 101
sites and $(b)$ is the analytical calculation. The step for $|k|>k_F$ disappears as soon as $k_F$ becomes larger than the first-Brillouin-zone vector $\pi/a$
or the correlation length, $\xi$ becomes small. The normal-metal DOS at the Fermi energy is $N_0=0.037$/eV, the Dynes broadening is
$\Gamma=0.01$ meV, and the gap is $\Delta=0.75$ meV.
                }
\label{Go}
\end{figure}

Expression Eq. (\ref{GBCSr}) lends itself to numerical implementation. This
can be interesting when trying to solve problems with spin chains
that require an all-numerical approach. We have computed Eq. (\ref{GBCSr})
by considering a finite 1-D array of sites on the superconductor
and compared with the results of the analytical calculation. The
agreement is very good even for rather small sets of the 1-D array of sites. Figure~\ref{Go}
shows the comparison of the density of states $-Im\, G_{BCS} (\vec{k}, \omega+i0^+) /\pi$
computed using both schemes. The  shown case is for $k_F=0.15\; a_0^{-1}$
and $a=3.36$ \AA\ that we have used to describe the $\beta$-Bi$_2$Pd superconductor~\cite{Choi_2018,Mier2021b}

The calculation of the density of states $-Im\, G_{BCS} (\vec{k}, \omega+i0^+) /\pi$
also reveals a cutoff for $k > k_F$ when $k_F$ is smaller than
the Brillouin zone value $\pi/a$. 
This reflects the fact that small values of $r$ are not well-taken care of
by the real-space Green's function. As a consequence, values $k > k_F$
will behave pathologically. However, the density of states becomes
well-behaved as soon as $k_F > \pi/a$, which corresponds to cases
of substantial band folding, or wide-band superconductors.
In the case of spin chains, large folding can be also obtained for
very diluted spin chains, where the distance between impurities
is much larger than the superconductor lattice. The
above procedure then works well in the limit of diluted spin chains \cite{Pientka2013,Paaske2016}.
The cutoff in density of states $-Im\, G_{BCS} (\vec{k}, \omega+i0^+) /\pi$
also disappears in the case of small correlation lengths. Showing
that there are two important length scales, $2 \pi/k_F$ and
the correlation length $\xi$. In the following calculations, we have always used the BCS value $\xi=\hbar v_F/\pi \Delta$ where $v_F$ is the free-electron Fermi velocity.

\section{In-gap-bands with Green's functions}
\label{Shiba}
In-gap or in-gap state bands are easily accounted for within
the approximation of classical spins\cite{Yu,Shiba,Rusinov}. We
adopt a spin $sd$ model, also known as Kondo Hamiltonian, that
separates the impurity action into charge and spin contributions
given by the potential scattering term $K_j$ and the exchange term $J_j$.
The Kondo Hamiltonian in the previous Nambu basis is given by
\begin{equation}
	\hat{H}_{impurity} = \sum_j^N (K_j \tau_z\sigma_0+J_j\vec{S_j}\cdot\vec{\alpha})\;
	\label{Kondo}
\end{equation}
where the sum over $j$ is over the impurities of the chain. The
atom spin is assumed to be classical and equal to $\vec{S_j}=(S_{j,x},S_{j,y},S_{j,z})$ {$= S(\sin{\theta_j}\cos{\phi_j}, \sin{\theta_j}\sin{\phi_j}, \cos{\theta_j})$}.
The electron spin is expressed via Pauli matrices in the Nambu basis set as:
$\vec{\alpha} = \frac{1 + \tau_z}{2}\vec{\sigma} + \frac{1 -
\tau_z}{2}\sigma_y\vec{\sigma}\sigma_y$, where $\vec{\sigma}$ is the spin
operator~\cite{Shiba}.

In the philosophy of the previous section, the effect
of the impurity chain can be  included using Dyson's equation. For
infinite periodic chains, this is done in reciprocal space, because
the equation becomes algebraic:
\begin{equation}
	G (\vec{k}, \omega) = G_{BCS} (\vec{k}, \omega) 
	+ G_{BCS} (\vec{k}, \omega)\Sigma (\vec{k}, \omega)G (\vec{k}, \omega).
	\label{Dyson}
\end{equation}
As above, the Green's functions, $G_{BCS}$ and $G$, and the self-energy $\Sigma$
are $4 \times 4$ matrices. The arithmetic involved to solve Dyson's equation
is just $4 \times 4$-matrix algebra.

Due to the locality of the Kondo Hamiltonian, Eq. (\ref{Kondo}),
and the mean-field character of the Bogoliubov-de Gennes theory,
the self-energy is easily computed. In 1-D and
assuming all impurities to be identical, it is simply
\begin{equation}
	\Sigma (\vec{k}, \omega)= \sum_{\vec{R}} \langle \vec{R}|
	\hat{H}_{impurity}| 0 \rangle e^{i \vec{k} \cdot \vec{R}} = K \tau_z\sigma_0+J\vec{S}\cdot\vec{\alpha},
	\label{Self}
\end{equation}
with $\vec{\alpha}$ defined above.
In the above expression, we have made used of the Bloch representation,
such that the matrix element $\langle \vec{R}|\hat{H}_{impurity}| 0 \rangle$
is only evaluated between unit cell $0$ and unit cell $\vec{R}$
of the periodic system.

%We can obtain direct information on the Shiba bands by plotting the
%density of states given by the imaginary part of the full retarded
%Green's function from Eq. (\ref{Dyson}),
%\begin{equation}
%\rho (\vec{k},\omega) = -\frac{1}{\pi} Im  [ 
%{G}^{1,1} (\vec{k},\omega) +{G}^{4,4} (\vec{k},-\omega)  ] 
%\label{rho}
%\end{equation}
%Where ${G}^{\nu,\mu}$ is  evaluated
%for the Nambu components $\nu$ and $\mu$.

\subsection{Evaluation of the effective Hamiltonian}
To compute the in-gap-bands is, however, not a simple task. As we saw in the preceding section, the superconducting Green's function, $G_{BCS}$, is not really
a Nambu resolvent. As a consequence, the usual method of diagonalizing
a Hamiltonian extracted from the Green's function~\cite{WangZhang},
%\begin{equation}
$	\hat{H} (\vec{k})=-G^{-1} (\vec{k}, \omega=0)$,
%	        \label{Dumb}
%\end{equation}
does not work. This cannot work because for $G^0$, this scheme gives something
approximate to a flat band for $-k_F < k < k_F$ at several hundreds
of meV depending on the electron density of the superconductor,
and adding an exchange coupling in the range of eV, just splits 
the band orders of magnitude away from the gap energy.

In order to solve this problem, we notice that we have
to generalize the resolvent equation. To do this,
we expand $G (\vec{k}, \omega)$ to first order in $\omega$, and
we identify this to the resolvent equation, Eq. (\ref{GoBCS}).
The resulting Hamiltonian comes from a renormalized Green's function
and shows the correct in-gap dependence:
\begin{equation}
	\hat{H} (\vec{k})=-\left (\frac{\partial G^{-1} (\vec{k}, \omega)}{\partial \omega} 
	\right )^{-1}_{\omega=0}G^{-1} (\vec{k}, \omega=0),
                \label{Hk}
\end{equation}
The results are excellent, the bands perfectly match the PDOS obtained
from the imaginary party of the retarded Nambu Green's function,
and all in-gap states properties are retrieved.
This is to be expected because the condition for
finding the bands or eigenvalues of Eq. (\ref{Hk}) is 
the condition of singularity for the Green's function for
small $\omega$. 

\subsection{Calculations in real space}

Here, we are going to compare with real-space calculations in order to 
describe the possible topological phases as well as in-gap states of other nature.
We assume we can express the electronic states in a local basis set, compact to the atomic sites,
that do not overlap and can be taken to be a tight-binding orthonormal basis set with
a total of $N$ orbitals or sites.

In this case, Dyson's equation is just a resolvent equation for a $4N\times4N$ matrix:
\begin{equation}
\label{Dyson2}
    \hat{G} = [\hat{G}_{BCS}^{-1} - \hat{H}_{I}]^{-1}
\end{equation}
Where $\hat{G}_{BCS}$ is the retarded Green's operator for the BCS Hamiltonian from Eq.(\ref{BCS}) and $\hat{H}_I = \hat{H}_{impurity} + \hat{H}_{Rashba}$. The $\hat{H}_{Rashba}$ includes the Rashba interaction as described in
the next section.

In this case, we evaluate the real-space density of states by projecting the
density of states on the tight-binding orbitals. This projected density
of states (PDOS) on orbital $i$ or spectral function is given by 
\begin{equation}
\rho (i,\omega) = -\frac{1}{\pi} Im  [ 
{G}^{1,1}_{i,i} (\omega) +{G}^{4,4}_{i,i} (-\omega)  ],
\label{rho_i}
\end{equation}
where ${G}^{\nu,\mu}_{ii}$ is the resulting Green's function evaluated on orbital $i$
for the Nambu components $\nu$ and $\mu$ by solving Dyson's equation. Thus, the calculations for finite chains are performed on a 2-D finite mesh of the 3-D superconductor, where a few sites without impurity interactions are left around the impurity chain. Our calculations are quite robust against
the number of free superconducting sites left around the impurity
chain, including subsurface layers, probably due to the 3-D character of the superconducting Green's function.

\subsection{Rashba self-energy}
In the same spirit as above, we can introduce the spin-orbit coupling for
a surface, using the self-energy for the Rashba Hamiltonian. In the
tight-binding
electron basis, the non-locality of the Rashba Hamiltonian makes it
formally similar to a nearest-neighbor hopping term,
\begin{eqnarray}
\label{Rashba}
    \hat{H}_{Rashba}&=&i \frac{\alpha_R}{2a}
	\sum_{i,j,\alpha, \beta}[\hat{c}^{\dagger}_{i+1,j, \alpha}
(\sigma_y)_{\alpha, \beta} \hat{c}_{i,j, \beta} \nonumber \\
&-&\hat{c}^\dagger_{i,j+1, \alpha}
(\sigma_x)_{\alpha, \beta} \hat{c}_{i,j, \beta} +h.c.]
\end{eqnarray}
where ${\alpha, \beta}$ are spin indexes. The lattice parameter of the substrate
is $a$, and the factor of $2 a$ comes from a finite-difference scheme to
obtain the above discretized version of the Rashba interaction.

Transforming to a 1-D reciprocal space and
using the Nambu basis set, the self energy becomes
\begin{equation}
\label{Self_Rashba}
\Sigma (k, \omega)= 
	2 \alpha_R \sin (k a) \tau_z \sigma_y
\end{equation}
For higher dimensions, we use the real space representation
given by the above Hamiltonian and we do the Fourier transform
to reciprocal space using a truncated unit-cell summation.

\section{Winding number and topological phase space for spin chains on a wide-band superconductor}
\label{IV}

The presented methodology based on Green's function permits us to compute both infinite and finite
spin chains on superconductors. We can easily put the bulk-boundary correspondence
principle\cite{Asboth,Kitaev_2001} to test as well as to characterize the topological superconducting phases
resulting from the in-gap states.

\subsection{Topological invariants}
Tewari and Sau \cite{Tewari2012} studied the topological properties of 1-D spin chains in one and two dimensions. They showed that the in-gap electronic structure induced by a spin chain on a 1-D superconductor leads to
phases compatible with the BDI class\cite{topo_calss}. This classification results from the chiral symmetry characterizing the system.

Due to the presence of magnetic interactions, time-reversal symmetry is broken in the model of a FM spin chain. However, the following antiunitary operator may be defined, $\mathcal{T} = \tau_0\sigma_0\mathcal{K}$, where $\mathcal{K}$ is the complex conjugate operator. The Hamiltonian satisfies $\mathcal{T}\Hat{H}(k)\mathcal{T}^{-1} = \Hat{H}(-k)$, this symmetry is the so-called \textit{generalized time reversal}\cite{heimes_2015} or \textit{spin-rotation time-reversal }\cite{Sato_2017} symmetry. Additionally, particle-hole symmetry is defined by the operator $\mathcal{P} = \tau_x\sigma_0\mathcal{K}$ and it is present on every BdG hamiltonian by construction. The combination of the two is the chiral symmetry and the corresponding operator is the product of the two previous ones, $\mathcal{C} = \mathcal{P}\mathcal{T} = \tau_x\sigma_0$. As a consequence, the Hamiltonian of the system can be written, in a rotated basis, under the form:
\begin{equation}
	H(k) = \begin{pmatrix}0 & A (k) \\ A^\dagger(-k) & 0\end{pmatrix}
		\label{chiral}
\end{equation}
Here $A$ is a $2\times 2$ matrix in the spin sector.
The representation of Eq. (\ref{chiral}) is easily obtained by changing the basis set from the Nambu basis expressed
in fermionic operators $\hat{c}$ and $\hat{c}^\dagger$ to a basis set expressed in terms of Majorana operators $(\hat{c}\pm\hat{c}^\dagger)/\sqrt{2}$. 

For a $2 \times 2$  Hamiltonian, Eq. (\ref{chiral}) can be written as:
\begin{equation}
	H(k) = d_x (k) \tau_x + d_y (k) \tau_y,
	\label{hd}
\end{equation}
where the change of basis has permitted us to have a zero component of $\tau_z$ because $\tau_z$
anticommutes with the Hamiltonian and defines the chiral symmetry of the system \cite{Asboth}. In the fermionic
basis set of the original Hamiltonian \cite{heimes_2015,Kobialka}, the symmetry representation is the above $\mathcal{C}=\tau_x\sigma_0$.
\begin{figure*}[t]
\begin{center}
\includegraphics[width=0.49\textwidth]{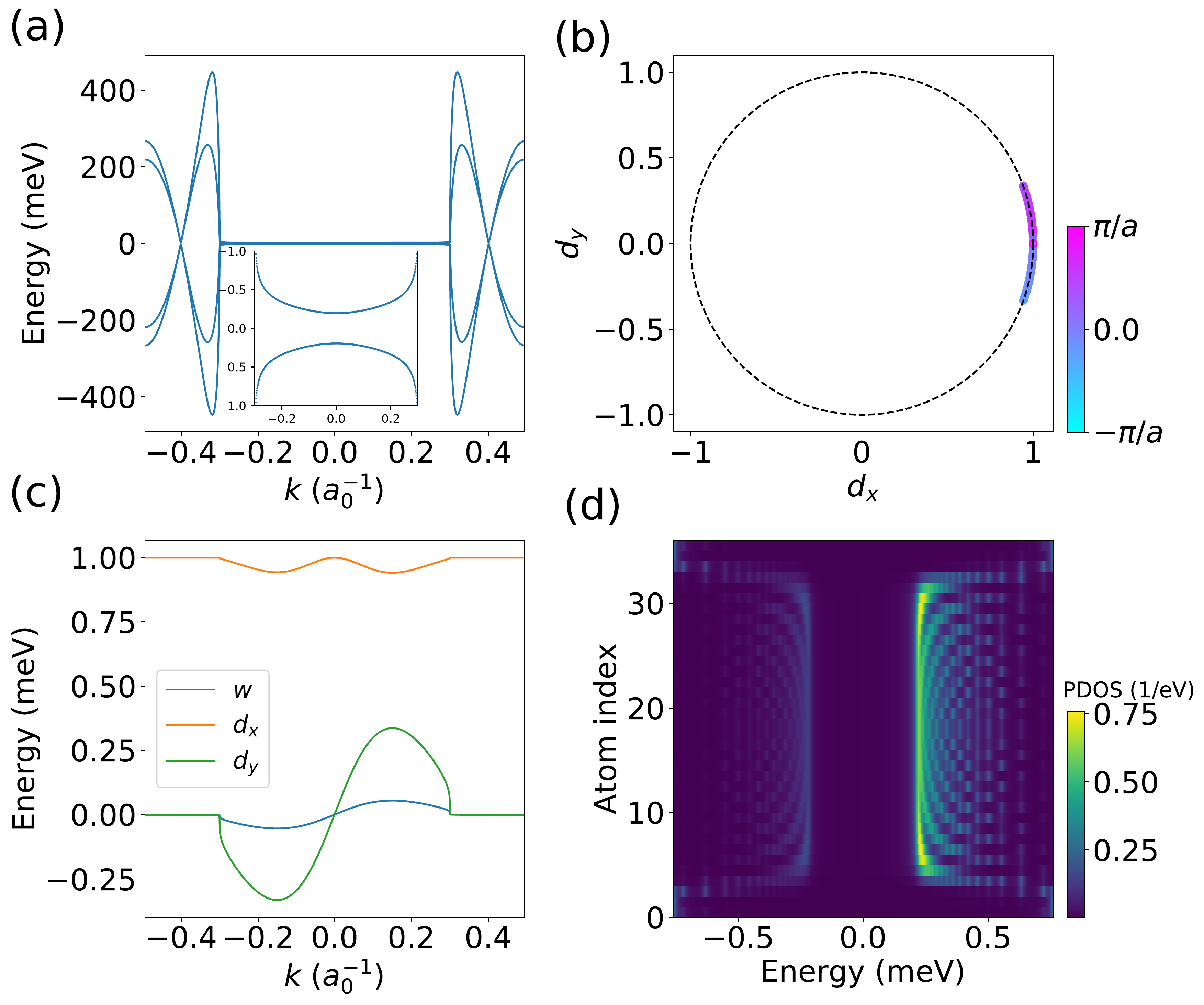}
\includegraphics[width=0.49\textwidth]{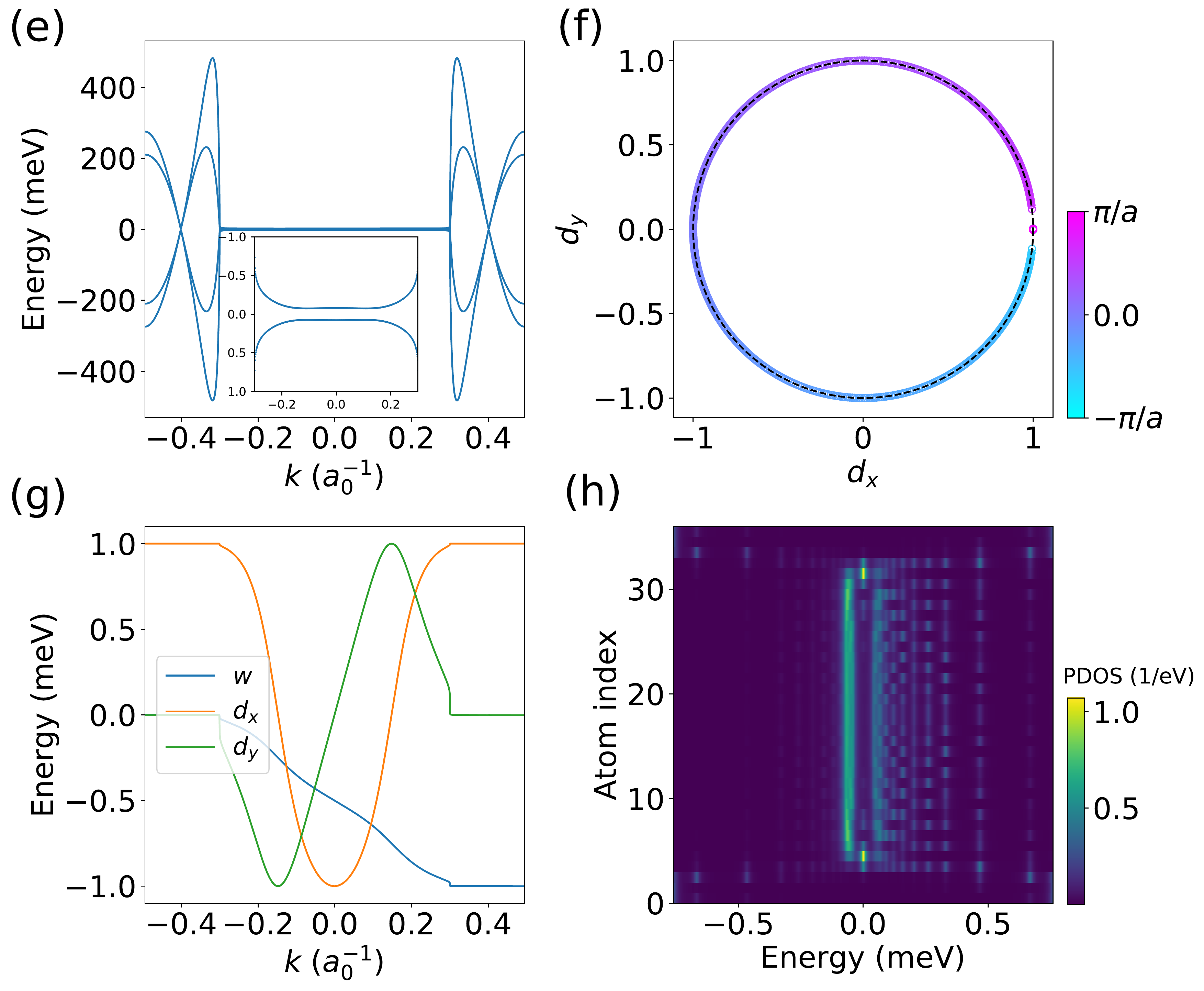}
\end{center}
        \vspace{-0.3cm}
	\caption{
		In-gap states for ferromagnetic spin chains in trivial ((a) to (d)) and topological ((e) to (h)) state. The Kondo coupling is $J=2.0$ eV (left) and $J=2.7$ eV (right), potential scattering $K=5.5$ eV and Rashba coupling $\alpha_R = 3.0$ eV-\AA, the Fermi vector is $k_F=0.3\; a_0^{-1}$ and the spin is  5/2 (like Cr or Mn), the metal density of states at the Fermi energy is $N_0=0.037$/eV and the superconducting gap is $\Delta = 0.75$ meV (like in $\beta$-Bi$_2$Pd [\onlinecite{Mier2021b}]). (a) and (e) The infinite spin chain has 4-bands with values close to the superconducting gap for values of $k$ between Fermi vectors, insets show the two lower bands where we can observe the trivial (a) and topological (e) gap. (b) and (f) plot the trajectory of the normalized $\vec{d} (k)$  as $k$ is varied. To facilitate the visualization of the trajectory, $k$ values go from $-\pi/a$ (in cyan) to $\pi/a$ (in magenta). In (b) all points remain in the vicinity of $(d_x, d_y) = (1,0)$ so no turn is completed about zero, on the other hand, in (f) we can observe a complete anticlockwise turn. (c) and (g) show the evolution of $d_x$ (orange), $d_y$ (green) and $w(k)$ (blue) as a function of $k$. In (c) we observe that $w$ has a final value of $w=0$ indicating that the system is a trivial state. In (g) $w(k)$ goes from 0 to $w=-1$, showing that here the system is a topological state. (d) and (h) show the result of a finite 30 atomic chain calculation with same parameters as the analytical one. We plot the PDOS in 2-D map as a function of the atomic site in the chain versus the energy. The spectra in (d) shows in-gap states at a lowest energy of $\sim 0.1$ meV, no zero-energy edge states are obtained here. In (h), we observe one zero energy state on each edge of the chain. The calculations for infinite chains are 1-D calculations, however, the finite-chain calculations
		correspond to finite chains on a 2-D finite mesh of the 3-D superconductor.
		}
	\label{winding}
\end{figure*}

The BDI class has a $\mathbb{Z}$ topological invariant. This is the winding number, $w$, related to the vector $\vec{d} = (d_x,d_y)$. As $k$ changes, $\vec{d}$ describes a closed trajectory. The winding number, $w$, is an integer that corresponds to the number of turns described by $\vec{d}$ about the origin. In order to change topological
phase, and change $w$, the superconductor gap has to close. 
This happens when the determinant of the Hamiltonian is zero. From Eq. (\ref{hd}), the
determinant of $H(k)$ will be zero when $\vec{d}$ is zero. 

However, as Tewari and Sau \cite{Tewari2012} emphasize, the Hamiltonian is a $4 \times 4$
matrix, and in order to keep the above description in the particle-hole sector (the $\tau$ matrices)
we identify the determinant of the $A$ matrices with the winding vector $Det (A (k) ) = d_x (k) - i d_y (k)$ in order to take into account when the determinant of the full Hamiltonian
becomes zero. From the $4\times 4$-Bogoliubov-de Gennes Hamiltonian, Eq. (\ref{BCS}) with the
addition of the impurity Hamiltonian, Eq. (\ref{Kondo}), and the Rashba term, Eq. (\ref{Rashba}),
we obtain:
\begin{eqnarray}
    &Det (A(k)) &= (H_{1,1} (k) +  H_{1,3} (k)) \times (H_{2,2} (k) +  H_{2,4} (k)) -\nonumber \\
    & &   (H_{1,2} (k) +  H_{1,4} (k)) \times (H_{2,1} (k) +  H_{2,3} (k)). 
\end{eqnarray}

Tewari and Sau \cite{Tewari2012} also show that a lower-symmetry class invariant can
be defined. This is the usual D-class $\mathbb{Z}_2$ invariant that is given by the parity of the winding number \cite{Tewari2012}. In 2-D superconductors, the Rashba interaction
leads to non-real matrix elements and the symmetry of
the 1-D system is reduced. However, we find
that even for 2-D substrates, the winding number still gives
results in agreement with the appearance of MBS in finite chains. It is then interesting to classify the topology of the spin-chain systems by their winding number, $w$.

The winding number is given by evaluating the number of turns of $\vec{d}$ about zero, given by the expression:
\begin{equation}
	w = \frac{1}{2 \pi} \int^{\pi/a}_{-\pi/a} d k (d_x \frac{d }{d k}d_y - d_y \frac{d}{d k}d_x),
	\label{wn}
\end{equation}
where $\vec{d}$ has been previously normalized.
Mathematically equivalent expressions can be obtained by using the trajectories in the complex plane
of $z=Det(A)/|Det(A)|$ as shown in Refs. [\onlinecite{Tewari2012}]  and [\onlinecite{Asboth}]. But they involve the evaluation
of the $log (z)$ that plagues the computation
with numerical problems due to artificial discontinuities caused by its branch cut.
Expression (\ref{wn}) however, is numerically simple and accurate to evaluate.

%Figure \ref{winding} shows two examples of trivial and non-trivial topological states, the %system is a ferromagnetic spin chain, with
%potential scattering of 5.5 eV, Rashba coupling of $\alpha_R = 3.0$ eV-\AA, for a spin of 5/2 %(like Cr or Mn), an electron density such that the Fermi vector is $k_F=0.3\; a_0^{-1}$ and %exchange
%or Kondo coupling of 2.0 eV and 2.7 eV, respectively.\\ \\

The $\mathbb{Z}_2$ topological invariant is calculated from the Pfaffian of the system. For a chiral Hamiltonian written as in Eq. (\ref{chiral}) the Pfaffian can be easily evaluated
using
   $Pf [H(k)] = Det [A(k)]$.
And the $\mathbb{Z}_2$ topological invariant, ${Q}$, becomes: 
\begin{eqnarray}
    {Q}& = & sign \,[ Pf [H(k = 0)]\times Pf [H(k = \pi/a)] ]\nonumber \\
    &=& sign \, [ d_x (k = 0)\times d_x (k=\pi/a)]
    \label{pfaffain}
\end{eqnarray}

\textcolor{black}{This equation shows that 
the Rashba Hamiltonian at $k=0$ and $k=\pi/a$ does not enter
in the determination of the above topological invariant since it is zero (see, e.g., Eq. (\ref{Self_Rashba})). For the same reason the} trajectories of $\vec{d}(k)$
wrap around zero only once, leading to winding
numbers that only take $-1$, $0$ or $+1$ values\cite{heimes_2015,Li_2018}. 

As an example, we calculate the in-gap bands and topological invariants for an infinite 1-D ferromagnetic chain on a superconductor. Panels (a) to (d) from Fig.~\ref{winding} correspond to a trivial state of the system. Fig.~\ref{winding} (a) shows the renormalization of the bands obtained from Eq. (\ref{Hk}) for $ -\pi/a \leq k \leq \pi/a $. As observed, the bands reach high energies for $k$ values, $|k| > k_F$, as the renormalization is not correct for $k$ in this range. The inset shows the two lower bands in for $-k_F \leq k \leq k_F$. Figure \ref{winding} (b) depicts the normalized trajectory described by the vector $\vec{d}$ in the complex plane. The $k$ points are labelled by a gradient of color going from $-\pi/a$ (in cyan) to $\pi/a$ (in magenta), in this case, $\vec{d}$ makes small oscillations around $(d_x, d_y) = (1, 0)$, meaning that the winding number in this case is $ w= 0$. On Fig.~\ref{winding} (c) we can follow this evolution: $d_x$ (orange curve) stays close to 1 and $d_y$ (green curve) describes a sinusoidal trajectory around 0 as we sweep $k$. The evolution takes place for $k$ values in the range $(-k_F, k_F)$, however, for $|k|>k_F$ $d_x$ is one and $d_y$ remains zero. Indicating that points for $|k|>k_F$ contribute trivially to the topology of the system. The blue curve depicts the evolution of the cumulative value of $w$:
\begin{equation}
	w (k) = \frac{1}{2 \pi} \int^{k}_{-\pi/a} d k' (d_x \frac{d }{d k'}d_y - d_y \frac{d}{d k'}d_x),
        \label{wnk}
\end{equation}

\begin{figure}[t]
\begin{center}
\includegraphics[width=0.48\textwidth]{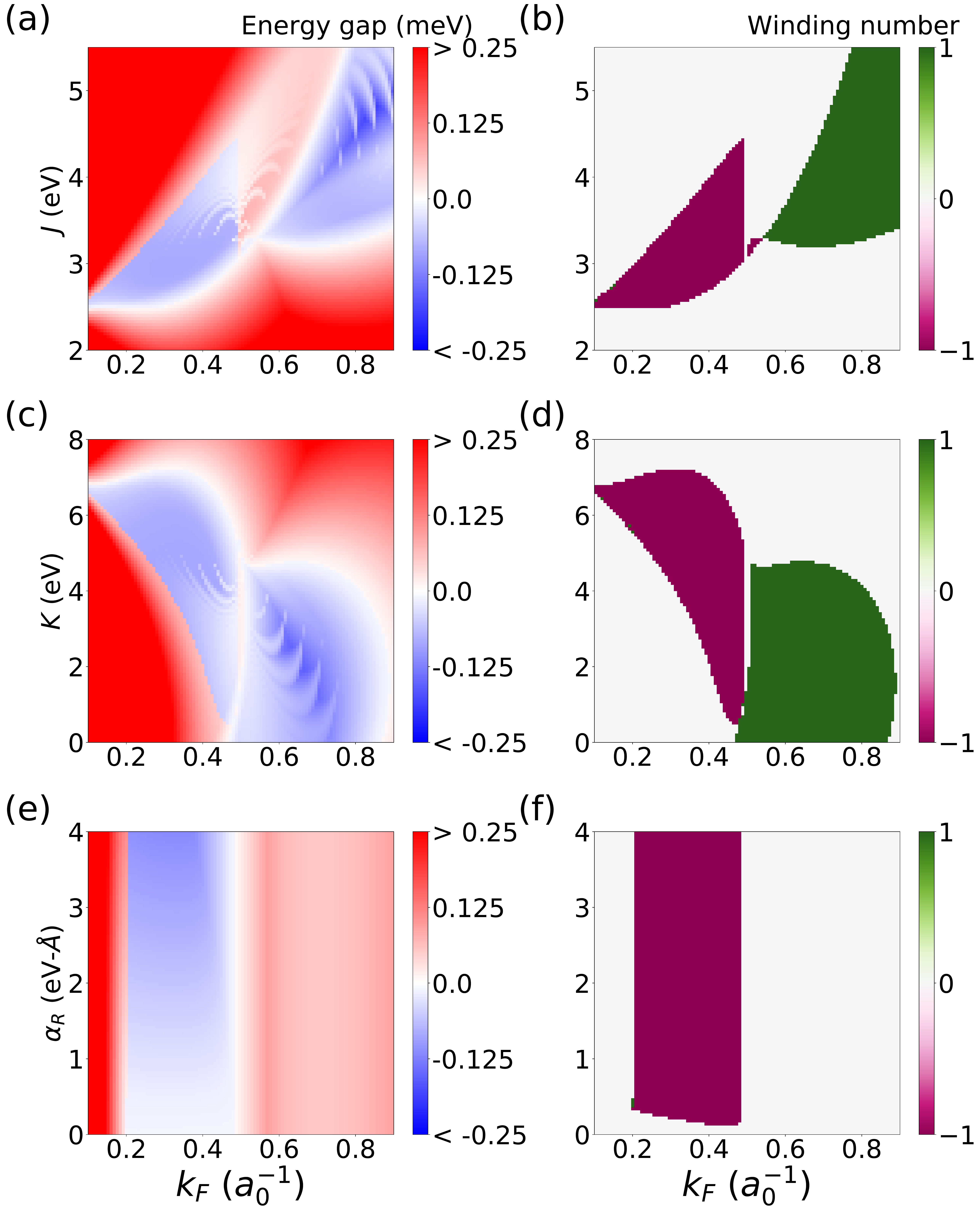}
\end{center}
        \vspace{-0.3cm}
	\caption{
		Phase diagrams obtained for a ferromagnetic spin chain with normal-metal DOS at the Fermi energy $N_0=0.037$/eV, $\Delta = 0.75$ meV and spin $s=5/2$. (a) and (b) Phase diagrams as a function of the Kondo coupling $J$ versus the Fermi wave vector of the system $k_F$, with potential scattering $K = 5.5$ eV and Rashba coupling $\alpha_R = 3.0$ eV-\AA. (a) Energy gap of the system multiplied by the $\mathbb{Z}_2$ topological invariant, $Q$, allowing for differentiation of trivial ($Q = 1$) and topological ($Q=-1$) phases. (b) Winding number, $w$ as a function of $J$ and $k_F$. The green areas correspond to $w=-1$ (cases like the one shown in Fig.~\ref{winding} (f)) and the magenta areas to $w=1$, here the winding vector, $\vec{d}$ completes a turn in the opposite direction. (c) and (d) Phase diagrams as a function of the potential scattering, $K$ versus $k_F$ with Kondo coupling $J=3.0$ eV and Rashba coupling $\alpha_R = 3.0$ eV-\AA. (e) and (f) Phase diagrams as a function of the Rashba coupling, $\alpha_R$ versus $k_F$ with Kondo coupling $J=3.0$ eV and potential scattering $K = 5.5$ eV.
		}
	\label{phase_diag}
\end{figure}
The $\mathbb{Z}_2$ invariant, $Q$, is calculated from Eq. (\ref{pfaffain}). For the present case of $k_F < \pi/a$, the points $k\geq k_F$ contribute trivially to the topological state of the system, such that we evaluate the  Pfaffian in Eq.(\ref{pfaffain}) at $k=k_F$ insteaf of  $k=\pi/a$. This is justified by the fact that beyond $k=k_F$ the free-electron-like states disperse rapidly away from the gap and they cannot alter the
topology of the in-gap bands. This is reflected
by the absence of states beyond $k_F$ in the superconductor as we show on Fig.~\ref{Go}. Numerically, we test that $d_y\approx 0$ in the k-points
where we evaluate $Q$. When $k_F > \pi/a$, we strictly apply Eq. (\ref{pfaffain}). In the case of
Fig. \ref{winding},we obtain $Q=1$ in good agreement with $w$. 

Figure \ref{winding} (d) shows a 2-D map of the PDOS calculated for a finite 30-atom chain with the same parameters as a function of the atomic site versus the energy. We calculate the PDOS on every site of the chain from Eq. (\ref{rho_i}). As we can observe, the in-gap states are distributed along the chain and the lowest energy states are found at $\sim 0.1$ meV. The absence of zero-energy edge states is in good agreement with the trivial state of the system. 

Panels (e) to (h) from Fig.~\ref{winding} correspond to a topological case. Here, we have increased the magnetic coupling to $J=2.7$ eV. The band structure has gone through a gap closing and the bands in Fig.~\ref{winding} (e) are topological. The trajectory of $\vec{d}$ completes a turn about zero, we can better observe the trajectory on Fig.~\ref{winding} (g), where $w(k)$ evolves from zero to -1. Again, the evolution takes place for $-k_F \leq k \leq k_F$ and the points in $|k|>k_F$ only contribute trivially. The calculation in a finite chain shows zero-energy edge states at both ends of the chain, as we expect from the bulk-boundary correspondence principle, Fig. \ref{winding} (h). 

The winding number, $w$, can be particularly difficult to evaluate because of the large number of k-points needed. The convergence depends on
the evolution of $\vec{d}$ with $k$. At $k=k_F$,
the band structure changes rapidly and so does $\vec{d}$. Large values of the
Rashba parameter, $\alpha$, lead to
smoother variations of $\vec{d}$, permitting
a more accurate evaluation of $w$ with fewer k-points. In the same way, the evaluation of gradients depends on the used discretization steps. It is
particularly critical to use small $\omega$ steps
for the evaluation of Eq. (\ref{Hk}) as well as a small
imaginary broadening for the Green's functions. The behavior of $d_y$ with $k$ is a stringent
test to check for the convergence of
the numerical calculations. Not only should $d_y$ equal zero at $k=0$ and $\pm\pi/a$, but it should be odd with $k$, as our results of Fig.~\ref{winding} show.

\subsection{Topological phase space}
By systematically evaluating the topological invariant ${Q}$ and the winding number on a parameter space, we can create phase diagrams that we will use to determine the topological state for any given parameters. Figure \ref{phase_diag} shows phase diagrams of a ferromagnetic atomic chain as a function of magnetic coupling $J$ versus $k_F$ (Fig. \ref{phase_diag} (a) and (b)), as a function of potential scattering $K$ versus $k_F$ (Fig. \ref{phase_diag} (c) and (d)) and as a function of Rashba coupling strength, $\alpha_R$ versus $k_F$ in Fig.\ref{phase_diag} (e) and (f).  The panels on the left row of Fig. \ref{phase_diag} depict the energy gap of the system multiplied by $Q$, like this, the topological phases are plotted as a negative gap (in blue) and in the trivial ones the gap is positive (in red). As expected, the topological phases corresponding to $w = +1, -1$ perfectly match  the $Q=-1$ areas. 

%. It is worth noting that a gap closing does not necessarily mean a TPT, as we can see, on Fig. \ref{phase_diag} (a), for $J\sim 4.0$ and $k_F \sim 0.6-0.7$. The gap goes to zero but the system remains in the trivial state ($Q = 1$).
At a TPT, the gap of the system goes to zero. On  Fig.~\ref{phase_diag} (a), we can easily observe two wide white branches corresponding to the gap closing at $k=0$ going from $k_F\sim0.1\; a^{-1}_0$ to $k_F\sim0.75\; a^{-1}_0$, and at $k=\pi/a$ at low values of $J$ for $k_F>0.5\; a^{-1}_0$. In other cases, however, the gap closing at a TPT can be difficult to observe. For example, in Fig.~\ref{phase_diag} (a) for Fermi vector values such that $k_F < 0.5\; a^{-1}_0$ and $J$ couplings going from $\sim 2.5$ eV to $\sim 4.5$ eV, the topological character changes, but we do not see a clear zero gap in this area. Here, the gap closes at a $k^*$ point close to $k_F$, but this transition is very abrupt requiring a high number of k-points and a fine tuning of the parameters to properly observe the gap closing. We have observed that the band structure highly depends on the number of $\omega$ and k-points, this can result in numerical artifacts in the energy gap maps. An example of this, is the stripped structure we can observe in Fig.~\ref{phase_diag} (a) for high values of $J$ and $k_F>0.6$ $a_0^{-1}$. When we look closely to the band structure for these values and use a sufficiently high number of $k$ points, we conclude that these white areas are an artifact of the non-converged bands. The phase diagrams we show on Fig.~\ref{phase_diag} are obtained using $N_k = 1001$ which is not sufficient to obtain clean maps, as we have observed, calculations with at least $N_k = 10001$ are required to remove these artifacts. As we have discussed in the previous section, a similar problem arises for the convergence of the winding number.

The strong dependence of the topological character on the exchange coupling $J$ is
natural given the necessary presence of an exchange interaction to have
in-gap states. However, the potential scattering term, given by matrix-element $K$ in Eq. (\ref{Kondo}), has an important effect on the topology of the bulk bands. In the localized-basis
set, this term appears as an on-site term, and it does the effect of a chemical potential.
It will shift the on-site energies of the superconducting sites, and hence has an
important influence on the topological phase, Fig. \ref{phase_diag} (c) and (d).

For $k_F$ values beyond the Brillouin-zone border, $\pi/a$, a stark change of topological phase
is found in Fig. \ref{phase_diag}. We have checked that this frontier is indeed
there and not some numerical artifact by testing the appearance of MBS in finite chains. The
topological regions for $k_F < \pi/a$ are characterized by a negative winding number.
For $k_F > \pi/a$ the winding number changes to $+1$. Thus, an interface between two superconductors of very different electron density, such that one has a $k_F < \pi/a$ ,
and the other one has $k_F > \pi/a$, a spin chain straddling the interface will have
a change of winding number of 2, and hence present two MBS at the interface. Alternatively to change the sign of the winding number, we change the sign of $\alpha_R$ because it changes the sign of $d_y$. The behavior of a ferromagnetic spin chain with Rashba coupling can be compared with the behavior of a helical non-collinear spin  chain\cite{Pientka2013}. Following this analogy, changing the sign of the coupling $\alpha_R$, would change the chirality of the spin helix. As a consequence, in a magnetic chain with a domain wall separating two different chirality chains, we also find the appearance of two MBS\cite{Ojanen, Ojanen_2013} at the domain wall.

Figure \ref{phase_diag} (e) and (f) show the phase diagrams as a function of the Rashba coupling versus $k_F$ for $J=3.0$ eV and $K=5.5$ eV. As we can observe, the topological phase is  independent of the Rashba parameter. However, the winding number phase diagram in Fig.\ref{phase_diag} (f) shows that the system is in the topological state only if we have a finite, non-zero $\alpha_R$, showing that Eq. (\ref{pfaffain}) should not be blindly applied, as topological phases on FM chains can only be achieved on systems with Rashba interaction, even if $\alpha_R$ is infinitesimally small\cite{heimes_2015, Brydon}. Moreover, in Fig.\ref{phase_diag} (e) we can see that the topological gap becomes bigger with an increasing $\alpha_R$, giving better protection to the MBS that arise in finite systems. Hence, the role of the Rashba interaction is to facilitate the triplet pairing, even though
the ferromagnetic ordering in the chain can suffice to locally drive the superconductor into the topological phase.

\section{Numerical studies of topological phases}
\label{V}
In the previous section we have shown that the topological phase can be determined for a ferromagnetic infinite chain. We now want to study the validity of the topological invariants in finite systems, in particular, in tens of atom chains on 2-D superconductors, which can be compared with experimental measurements\cite{Mier2021b}. We create a 2-D superconducting array, without loss of generality, the magnetic impurities are located along the $\vec{x}$ direction in an atomic chain, all spins are oriented perpendicular to the substrate along the $\vec{z}$ direction, creating a ferromagnetically-ordered chain. We solve Dyson's equation, Eq.~(\ref{Dyson}), and the PDOS is calculated on every site using Eq.~(\ref{rho_i}). %The previous calculations with infinite ferromagnetic chains will allow us to identify the topological state for each case.
\begin{figure*}[t]
\begin{center}
\includegraphics[width=0.85\textwidth]{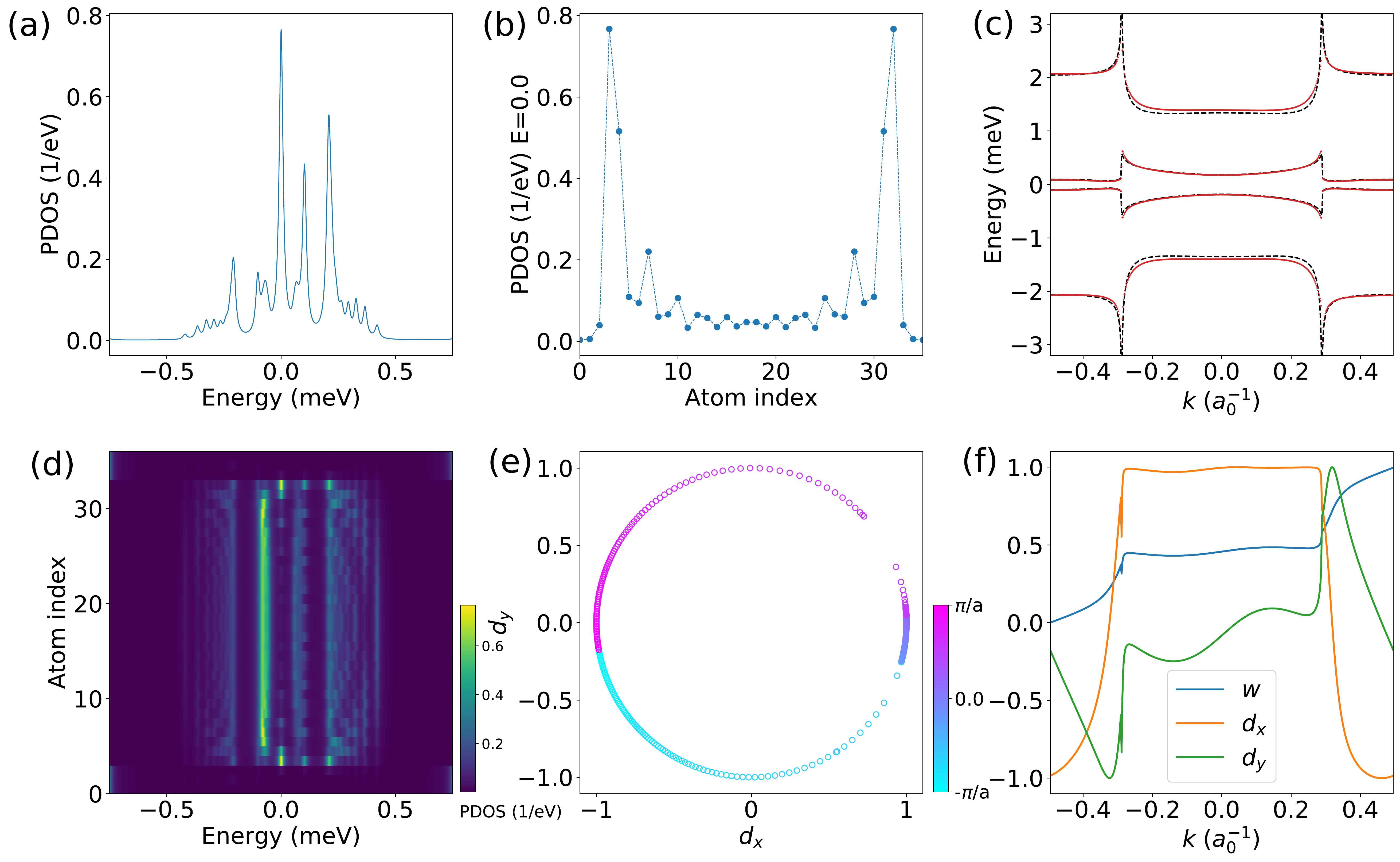}
\end{center}
        \vspace{-0.3cm}
	\caption{
	Numerical results of 30-atom impurity chain of a 2-D superconductor with dimensions $N_x=36$ and $N_y=5$. (a) Spectrum obtained on the first atom of the chain. (b) PDOS at zero energy along the chain's axis. (c) Renormalized bands, analytical calculation (black dashed lines) and numerical result (in red) for a 1001-atom chain in a 1-D superconductor. (d) PDOS spectra along the 30 atom chain. (e) Corresponding trajectory of winding vector $\vec{d}$, the color bar shows the $k$-points where $\vec{d}(k)$ is evaluated. (f) $d_x$, $d_y$ and cumulative winding number $w(k)$, Eq. (\ref{wnk}) as a fubction of $k$. Parameters: $\Delta = 0.75$ meV, $N_0=0.037$/eV, $k_F = 0.7\; a_0^{-1}$, $\alpha = 3.0$ eV-\AA\, $J=3.5$ eV, $K=5.5$ eV.
	}
	\label{chain}
\end{figure*}

We have verified that the in-gap states are not drastically affected by the change in dimensionality. By performing calculations on 2-D superconductors, we were able to observe that the extension of the in-gap states decays in about 5 sites in the perpendicular direction to the chain. The overall PDOS obtained along the chain and the in-gap states dispersion are largely unaffected by the change from 1-D to 2-D. In the case of a 3-D system, 3 layers are enough for the states to decay. On the present work, the calculations on finite chains are performed on 2-D superconducting arrays. However, for calculations of in-gap bands and topological invariants, a big number of atoms is required in order to attain the infinite-chain behavior, hence we limit ourselves to 1-D systems in order to reduce the computational time.

\subsection{Comparison with analytical calculations}

On Fig.~\ref{chain} we show the results for a finite 30-atom chain located at the center a 2-D rectangular superconducting array with dimensions $N_x = 36$ and $N_y=5$ sites. The exchange coupling is $J=3.5$ eV, the potential scattering ampitude is $K=5.5$ eV, Rashba coupling is $\alpha_R = 3.0$ eV-{\AA} and the Fermi vector is $k_F=0.7a_0^{-1}$, by looking at Fig.~\ref{phase_diag} (a) these parameters yield a topological solution with winding number $w=1$. On Fig.~\ref{chain} (a) we depict the spectrum obtained on the first atom of the chain, here a very pronounced peak can be observed at zero energy. On panel (b), we show the distribution of the PDOS at zero energy along the $\vec{x}$ axis, revealing that the zero-energy state is well localized at the ends of the chain. On Fig.~\ref{chain} (d) we show a 2-D map of the spectra obtained on every atom along the chain's axis, we can again note the presence of zero-energy edge states, whereas inside the chain we observe a finite energy gap. All of these features are in good agreement with the presence of MBS. As discussed in the previous section, the topological state of a given system can be determined from the study of the topological invariants.

We can calculate $G(\vec{k}, \omega)$ from the real-space Green's function $G_{i,j}(\omega)$ by using a finite Fourier transform, and using a sufficiently high number of atoms in a 1-D finite system. We then calculate the k-resolved Hamiltonian from the renormalized Green's function, Eq. (\ref{Hk}). Figure~\ref{chain} (c) depicts the numerically calculated bands (in red) for a 1001-atom chain with the same parameters as for the 30-atom chain. We plot the infinite-chain bands from the previous section as black dashed lines, showing good agreement with the numerical calculations. We show the trajectory of the vector $\vec{d}$ in Fig. \ref{chain} (e), making a complete turn about zero in the positive sense, resulting in $w=1$ and demonstrating the topological nature of the edge states obtained in the 30-atom chain, Fig. \ref{chain} (d). In contrast to the infinite-chain results, the winding number determined by $d_x$ and $d_y$ show some incorrect asymmetry with $k$, Fig. \ref{chain} (f), this asymmetry can be reduced by taking sufficiently small $\omega$ steps that improves the numerical precision of the derivative in Eq.~(\ref{Hk}). Also, small oscillations can appear in these curves due to the Fourier transform from the finite-chain in real space to $k$-space, Fig. \ref{chain} (f). In order to improve the results, a sufficiently high number of k-points and high number of atoms are required. The Dynes parameter, $\Gamma$, needs to be adjusted for better accuracy. Overall, these results show good agreement between finite and infinite chain calculations that is of special interest, because it shows that the topological state of a given system can be determined from strictly numerical calculations in finite systems.

\subsection{Numerical phase space}

In contrast to the infinite-chain analytical calculations of previous sections, finite-chain calculation has the advantage that the presence of MBS can be quickly discerned in a calculation. Moreover, the phase space can be explored by computing the in-gap electronic states projected on the first site of the chain. In the presence of MBS, zero-energy states will appear as parameters change.

\begin{figure*}[t]
\begin{center}
\includegraphics[width=0.9\textwidth]{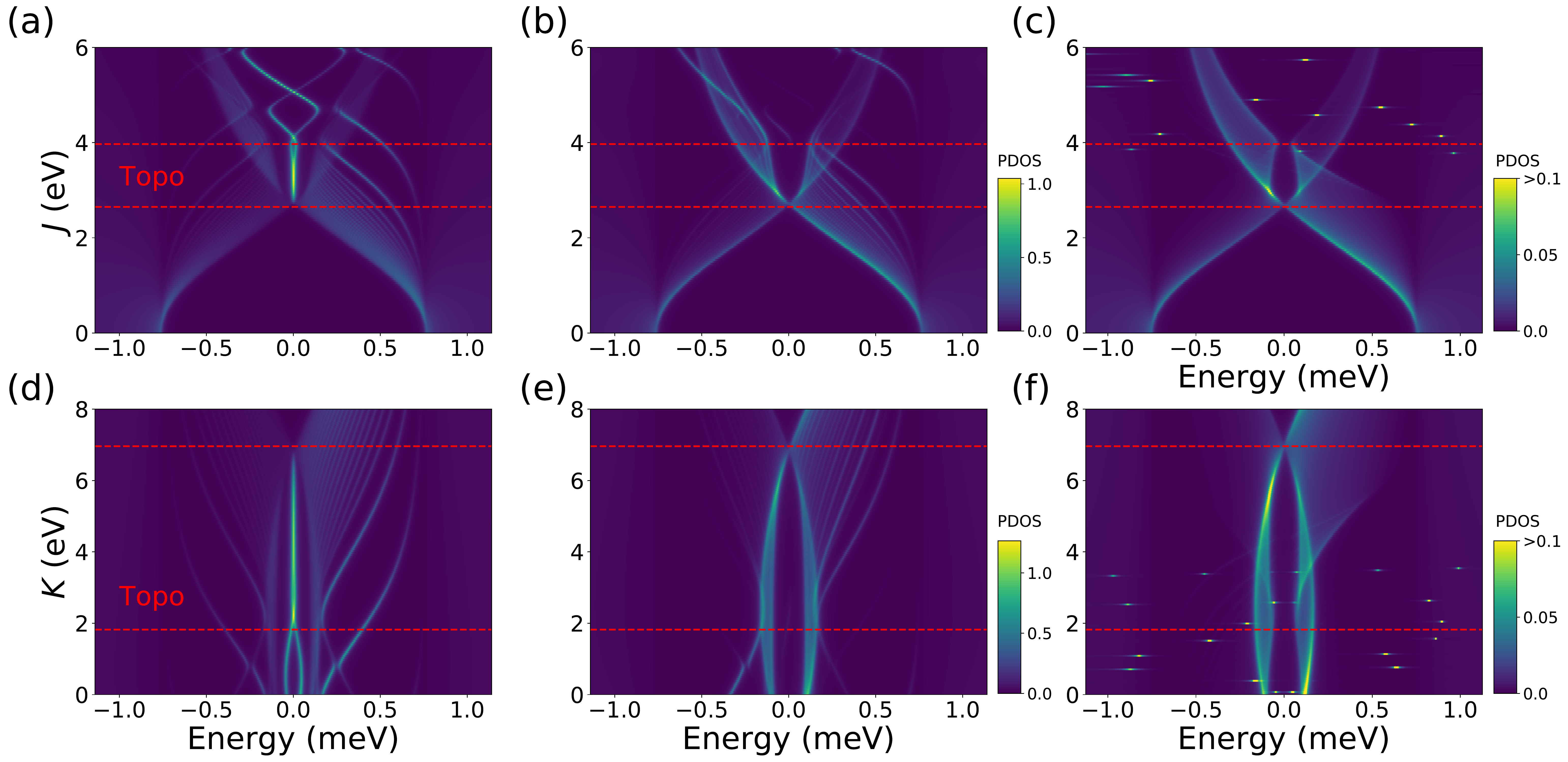}
\end{center}
        \vspace{-0.3cm}
	\caption{Evolution of in-gap states at the edges and at the center of the chain as a function of the couplings for exchange $J$ ((a) and (b) respectively) and potential $K$ ((d) and (e)) interactions. For comparison, the states at the center of an infinite chain from an analytical calculation are also shown in (c) and (f). As
    expected, the agreement between the spectra at the mid-site of the finite chain (b) and the site
    of the infinite chain (c) is excellent, as well as between (e) and (f). The main differences are due to long-range edge states that are absent from the infinite chain. Red dashed lines indicate the TPT as obtained from phase diagrams in Fig.~\ref{phase_diag}. Topologically non-trivial phases are found between the two horizontal lines. Parameters: $\Delta = 0.75$ meV, $N_0=0.037$/eV, $k_F = 0.4\; a_0^{-1}$, $\alpha = 3.0$ eV-\AA, the finite chains ((a), (b), (d) and (e)) are 30-atom long. The PDOS is in (1/eV) units.
	}
	\label{J_U_study}
\end{figure*}

To study the evolution of the edge states in the finite chains as we go through the TPT, we calculate finite 30-atom chains as a function of the parameters $J$ and $K$. In order to reveal the features proper to the edge of the chain, we compare the electronic structure as a function of energy for edge sites with the one at the center of the chain. Figure \ref{J_U_study} depicts the evolution of the edge states and the states at the center of the chain as the exchange ((a) and (b) respectively) and potential ((d) and (e)) couplings are varied. For comparison, we perform a calculation from the analytical solution $G (k,\omega)$ of an infinite chain, and we Fourier transform to real space, such that a site in an infinite chain can be evaluated ((c) and (f)). As expected, the agreement between Fig. \ref{J_U_study} (b) and (c) is excellent, as well as between (e) and (f). There are however some differences, particularly from states that cross the gap as the interactions change. These states are not present in the infinite-chain calculation and can be traced back to the projections on the edge sites, Fig. \ref{J_U_study} (a) and (d), showing that they are edge states extending into the center of the chain.

The red dashed lines indicate the TPT as found from the phase diagrams in Fig. \ref{phase_diag}. In good agreement, we find that MBS develop in (a) and (d) for the values of the couplings corresponding to topological phases. Moreover, the states that cross rapidly the Fermi energy when the couplings are changed can be determined to have no topological origin by comparison with Fig. \ref{phase_diag}.

A closer look to Fig.~\ref{J_U_study} (a) reveals that for higher values of $J$ in the topological state, the zero energy edge states begin to split. This is due to the finite size of the chain, Fig. \ref{J_U_study} corresponds to calculations with a 30-atom chain. For an increasing number of atoms, the splitting of the zero-energy peak occurs closer to the TPT, marked by the red dashed line. The TPT is marked by a gap closing of the bulk hamiltonian revealed by the crossing at $J\sim 2.7$ eV of the zero-energy in-gap states, Fig. \ref{J_U_study} (b) and (c). For the second transition at $J\sim 4.0$ eV, we observe a narrowing of the gap, but the gap closing is difficult to observe because a high number of k-points and $J$ values is required to observe this gap closing. A similar situation happens when tuning the potential scattering, $K$ in Fig. \ref{J_U_study} (e) and (f).

\subsection{Finite-chain spectral dependence on the number of atoms}

\begin{figure*}[ht]
\begin{center}
\includegraphics[width=0.9\textwidth]{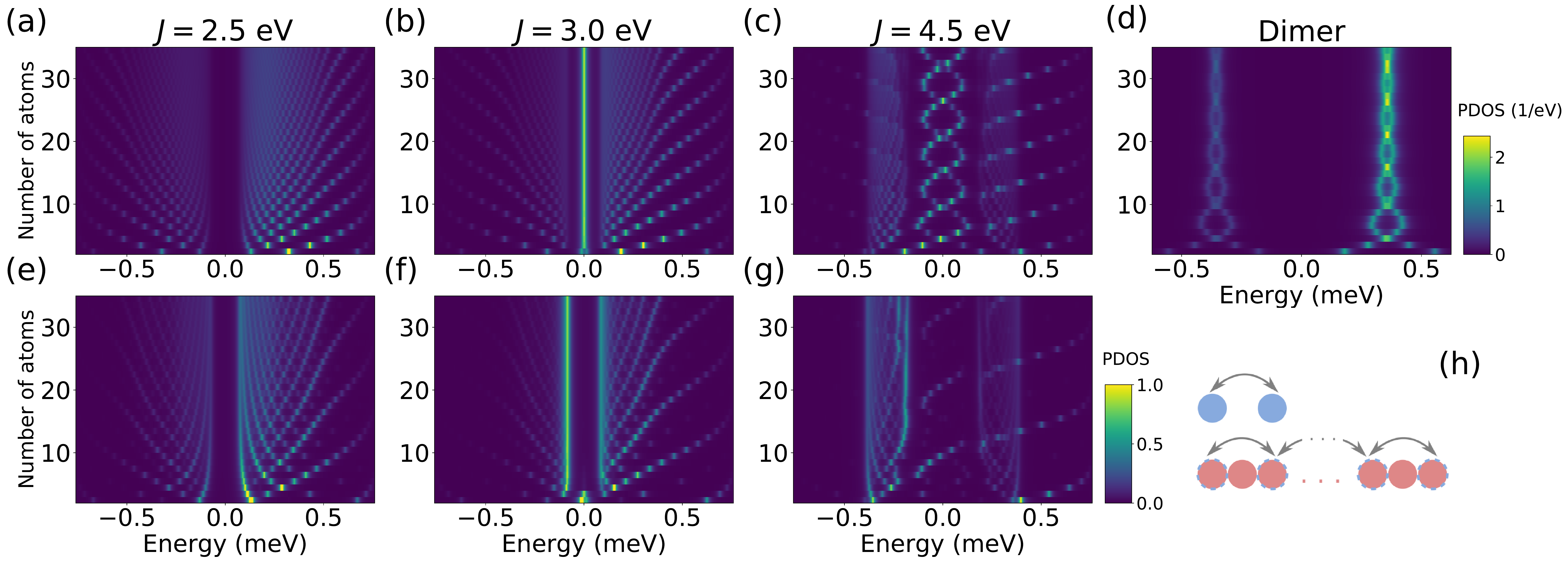}
\end{center}
        \vspace{-0.3cm}
	\caption{
    (a) to (c) ((e) to (g)) Evolution of the edge (center) states as a function of the number of atoms in the chain and for different Kondo couplings. For $J=2.5$ eV the lower energy in-gap states are at $\sim 0.1$ meV and well distributed between edge (a) and center of the chain (e). For a coupling of $J=3.0$ eV we observe a robust zero energy edge state for chains as short as 5 atoms (b) while the bulk spectra shows an energy gap. For $J=4.5$ eV the edge states oscillate around zero with a period of 5 atoms (c) the oscillations are also observable in the middle of the chain (g). Parameters: $\Delta = 0.75$ meV, $N_0=0.037$/eV, $k_F = 0.4\; a_0^{-1}$, $\alpha = 3.0$ eV-\AA\, $K=5.5$ eV. (d) Evolution of the ingap states in a dimer while varying the distance between the two magnetic impurities. The four in-gap-states oscillates with the same period observed in the chain. Parameters: Same as the chain but $J=3.2$ eV. (h) Scheme depicting the interaction between pairs of atoms in the chain (in red) and in the dimer (blue) here the periodicty of the interaction is set to 2 atoms for simplicity. The PDOS is in (1/eV) units.
	}
	\label{N_study}
\end{figure*}

The study of the spin state\cite{Mahdi2020,Mier2021b} of the chain while varying the magnetic coupling supports the occurrence of a topological phase transition at $J \sim 2.7$ eV (with parameters $U=5.5$ eV, $k_F=0.4$ $a_0^{-1}$ and $\alpha = 3.0$ eV-\AA) and, hence, the presence of MBS in this case. This is also in agreement with the phase diagram from Fig. \ref{phase_diag} (a), for $k_F=0.4\; a_0^{-1}$, the energy gap goes to zero at about $J=2.7$ eV, and the new gap changes character from trivial to topological. To further study these finite system states, we follow the evolution of the edge states while changing the number of atoms in the chain.

MBS are expected to be easier to detect as the chain length increases \cite{Mier2021b,Peng_2015} because the spatial overlap of their wave functions decreases. On Fig.~\ref{N_study} we show the evolution of the spectra on the first (top row) and middle atom (bottom row) of the chain as a function of number of atoms and for different $J$ coupling values. On panels (a) and (e) $J=2.5$ eV, the topological state has not been reached and the in-gap states are still far from zero energy. In the middle plots (panels (b) and (f)), we have increased the magnetic coupling to $J=3.0$ eV, this is after the system has undergone the TPT. On panel (b) we observe a robust zero energy state for chains as short as 5 atom-long. As the chain length increases, the edge state stays at zero energy. If we look at the spectra on the middle of the chain (Fig.~\ref{N_study} (f)), we can observe an energy gap, showing that the zero-energy state is well localized at the chain edges. The phase diagram of Fig.~\ref{phase_diag} (a) shows that for $J=3.0$ eV and $k_F=0.4\; a^{-1}_0$, the system is, indeed, in a topological state.

On panels (c) and (g) from Fig.~\ref{N_study}, the exchange coupling is $J=4.5$ eV and the spectra on the upper panel display an edge state with an oscillatory behavior around zero energy with a period of 5 atoms. On panel (g), we see that some of these edge states are extended inside of the chain. Oscillations of in-gap states has been reported by recent studies\cite{Schneider1}, suggesting that even for topological solutions, the MBS can interact and move away from zero energy. To better understand the nature of the oscillations, we look at the phase diagram on Fig.~\ref{phase_diag} (a). For these parameters the system is in the trivial state. Despite the edge states crossing at zero energy periodically, they are no-topological in-gap states.

Figure \ref{N_study} (d) depicts the in-gap states of a dimer of magnetic atoms in a superconductor as a function of their interatomic distance. On the $y$ axis we vary the distance between the two atoms. Four in-gap-states results from the hybridization of the FM dimer\cite{Choi_2018}. As the distance changes we observe an oscillatory behavior of the states. %The period of the oscillations is $\sim 5$ atoms, this periodicity highly depends on the Fermi-vector, $k_F$, and it coincides with the behavior we observe on the atomic chains: For smaller $k_F$ the period is reduced and for bigger $k_F$ the oscillations become bigger, this dependency is found in both chain and dimer. 
This points to a coupling between atomic pairs carried by RKKY interaction\cite{Rebola,Samir}. %The oscillations result from the $\sin{k_Fr}$ term in $G_{BCS}(r, \omega)$ Green's function. 
In the case of the dimer, the amplitude of the oscillations decays with the distance because the coupling between the two atoms becomes smaller as the two impurities move away. For very large interatomic distances, the dimer spectra tend to the spectra of a single impurity. However, in the case of the atomic chain, because we keep adding atoms, the coupling between pairs at a given distance is always present so the oscillation amplitude does not decay, a scheme of these interactions is depicted on Fig.~\ref{N_study} (h). %Moreover, if these edge states corresponded to MBS, we would expect the oscillations to reduce their amplitude as we add atoms to the chain\cite{Schneider2}, as the coupling between the Majorana is reduced. However, we have calculated chains up to 70 atoms and the oscillations keep the same amplitude.

For different parameters, we also find oscillatory behavior about zero energy in the topological phase when the exchange coupling is very large. In this case, the interactions between the edge MBS are
not negligible and we reproduce the same behavior as the one reported in Ref. [\onlinecite{Schneider2}]. In order to obtain topological or trivial oscillations, we find that the exchange coupling, $J$, needs to be large enough to induce the oscillatory behavior of the in-gap states as the number of atoms is increased.

The atomic manipulation capabilities of the STM allows us to study the evolution of the in-gap structure as atoms are added to the chain\cite{Mier2021b, Schneider1, Schneider2}. Hence, the above real-space studies permit us a direct comparison with experiments.

\section{Summary and conclusions}
In this paper we have developed a theoretical framework using a Green’s functions approach to model wide-band superconductors  that correctly describe the band structure at the superconducting-gap energy range, despite of the large mismatch between the normal-metal band width of the superconductor and the pairing energy, $\Delta$. %From the Fourier Transform of the $G_{BCS}$ and the renormalization of the bands in Eq. (\ref{Hk}), we have obtained the in-gap bands describing an infinite ferromagnetic impurity chain in a superconductor.

The in-gap-bands obtained from the bulk Hamiltonian allows us to calculate the winding number and the $\mathbb{Z}_2$ topological invariant that determine the topological state of the system. We have thus computed phase diagrams that help us to easily classify our systems. Both the %Kondo coupling ($J$) and potential scattering ($K$) 
exchange and the potential-scattering interactions
can drive the system in and out of the topological phases. According to our calculations, for infinitesimally small Rashba couplings, %($\alpha_R$) values 
the topological phases can be accessed. It is worth noting the the convergence of the band structure and the topological invariants is not trivial, as a sufficiently high number of $\omega$ and $k$ points is generally required.% for certain sets of parameters. This may lead to a incorrect evaluation of the topological phases of the system.}

Our numerical calculations of finite systems show good agreement with the infinite chain, from which we could determine the topological state of edge states. This is of special interest because it shows that the topological state of the system can be evaluated from the finite system alone. This further open the possibility of using this methodology using Green's function that are derived for free-electron metals.

We have performed calculations of the topological properties of finite spin chains and their dependence on the number of atoms of the chain. We find recurring in-gap state oscillations about zero energy as the number of atoms is increased. These oscillations originate in large impurity-electron exchange couplings, but their topological behavior actually depends on the electronic density, fixed by the Fermi wave vector, $k_F$, in our studies.%With this information, we were able to distinguish trivial zero energy states that show an oscillatory behavior around the Fermi energy as we increase the number of atoms in the chain. The comparison with the results obtained for a magnetic dimer, point to a interaction between pairs of atoms rather that coupling between MBS.}

In summary, the present model is a promising tool that has already successfully described experiments on magnetic atoms manipulated with STM\cite{Mier2021b}. And will be of special interest in the search of topological phases of spin chains on s-wave superconductors.

\section*{Acknowledgements}
Financial support from the Spanish MICINN (projects RTI2018-097895-B-C44
and Excelencia EUR2020-112116) and Eusko Jaurlaritza (project
PIBA\_2020\_1\_0017) is gratefully acknowledged.

%\newpage
\bibliography{references}
\end{document}